\documentclass[aps,superscriptaddress,twocolumn]{revtex4-1}
\usepackage{amsmath}    % need for subequations
\usepackage{graphicx}   % need for figures
\usepackage{verbatim}   % useful for program listings
\usepackage{xcolor}      % use if color is used in text
\usepackage{subfigure}  % use for side-by-side figures
\usepackage{hyperref}   % use for hypertext links, including those to external documents
\usepackage{natbib}
\graphicspath{{./figures/}}
\usepackage{float}
\newcommand{\f}[1]{\textit{#1}}
\newcommand{\bo}[1]{\textbf{#1}}

\begin{document}

%\title{Effect of Ge-substitution on Magnetic Properties in the itinerant chiral magnet MnSi
\title{Effect of Ge-substitution on Magnetic Properties in the Itinerant Chiral Magnet MnSi}
\author{Seno Aji}
\email{senji77@dc.tohoku.ac.jp}
\author{Hidesato Ishida}
\author{Daisuke Okuyama}
\author{Kazuhiro Nawa}
\affiliation{Institute of Multidisciplinary Research for Advanced Materials, Tohoku University, Sendai 980-8577, Japan}
\author{Tao Hong}
\affiliation{Neutron Scattering Division, Oak Ridge National Laboratory, Oak Ridge, Tennessee 37831, USA}
\author{Taku J Sato}
\affiliation{Institute of Multidisciplinary Research for Advanced Materials, Tohoku University, Sendai 980-8577, Japan}
%\date{12 January 2019}

\begin{abstract}
We have investigated the effect of Ge-substitution to the magnetic ordering in the B20 itinerant chiral magnet MnSi prepared by melting and annealing under ambient pressure. From metallurgical survey, the solubility limit of Ge was found to be $x=0.144(5)$ with annealing temperature $T_\mathrm{an} = 1073$ K. Magnetization measurements on MnSi$_{1-x}$Ge$_x$ samples show that the helical ordering temperature $T_{\mathrm{c}}$ increases rapidly in the low-$x$ range, whereas it becomes saturated at higher concentration $x>~0.1$. The Ge substitution also increases both the saturation magnetization $M_\mathrm{s}$ and the critical field to the fully polarized state $H_\mathrm{c2}$. In contrast to the saturation behavior of $T_\mathrm{c}$, those parameters increase linearly up to the highest Ge concentration investigated. In the temperature-magnetic field phase diagram, we found enlargement of the skyrmion phase region for large $x$ samples. We, furthermore, observed the non-linear behavior of helical modulation vector $k$ as a function of Ge concentration, which can be described qualitatively using the mean field approximation.

\end{abstract}

\maketitle

\section{Introduction}

The B20 chiral compound MnSi has attracted continuous interest for decades \cite{jeong,thessieu,nakajima,tite,neubauer,bauer}. MnSi has a cubic chiral crystal structure with the noncentrosymmetric space group $P$2$_1$3. Due to the lack of inversion symmetry at the center of the magnetic Mn-Mn bond, the antisymmetric spin-spin interaction, called Dzyalonshinskii-Moriya (DM) interaction, becomes active.  This antisymmetric interaction introduces long-period modulation to the otherwise collinear ferromagnetic structure stabilized by the dominant ferromagnetic interaction. The resulting long-period helically modulated structure is established below $T_{\mathrm{c}}\simeq 30~\mathrm{K}$ under zero external magnetic field \cite{williams}. The modulation has 180 \r{A} periodicity, propagating along the [111] crystallographic axis \cite{ishikawa}. Recently, intriguing magnetic-skyrmion phase was found in MnSi under finite external field ($\sim 2000$ Oe) close to $T_{\mathrm{c}}$ \cite{muhlbauer}. The magnetic skyrmion is the swirling spin texture characterized by a topologically nontrivial skyrmion number. Because of the topologically protected nature of skyrmions, scrutiny on this itinerant chiral magnet has been drastically accelerated recently. At higher magnetic field ($H>5500$~Oe), a further transition takes place to the trivial fully polarized (induced ferromagnetic) phase.

One of interesting features in MnSi is that the helical ordering temperature $T_{\mathrm{c}}$ can be tuned by introducing chemical substitution of either Mn or Si site with the elements with different atomic radius and/or electron concentration. Attempts to change the helical ordering temperature of MnSi by chemical substitution of Mn by Fe or Co have resulted in a decrease of the helical ordering temperature \cite{grigoriev01,bauer2,dhital2}. Note that the Fe or Co substitution introduces both the positive chemical pressure and electron doping simultaneously. Under negative chemical pressure introduced by the isovalent substitution of Si with Ge, increasing helical ordering temperature has also been reported in several experiments, however in a contradicting way: In Ref.~\cite{sivakumar}, slight increase of $T_{\mathrm{c}}$ (less than 3 K) was reported for MnSi with 10\% Si substituted by Ge. In contrast, the other experiment shows that the helical ordering temperature can reach 39 K by replacing only 1\% Si with Ge \cite{nadya}. Quite recently, there appears a report on magnetic properties of MnSi$_{1-x}$Ge$_x$ in a wide $x$ range $0 \leq x\leq 1$, using high-pressure synthesis \cite{fujishiro}. As this work focuses on the topological transitions in the wide $x$-range, there are only one or two data points in $x<0.2$, and hence it is still unclear how magnetic properties vary in the low-$x$ range. 

In the present work, we undertook the detailed investigation on the effect of the Ge-substitution to the magnetic ordering in MnSi in the low $x$-range ($x<0.15$). We have performed metallurgical survey for the MnSi$_{1-x}$Ge$_x$ alloys with different nominal Ge concentrations and annealing temperatures, in order to clarify solubility limit of Ge in MnSi for ambient pressure synthesis. Utilizing the metallurgical information, we have successfully prepared polycrystalline samples of various Ge concentrations up to $x\sim ~0.15$. The magnetic properties of MnSi$_{1-x}$Ge$_x$ alloys have been investigated as a function of the Ge-concentration $x$. The helical ordering temperature $T_\mathrm{c}$ increases rapidly in the low-$x$ range ($x<0.1$), whereas it becomes saturated for $x>0.1$. In contrast, the magnetization measurements at the base temperature show the saturation magnetization and the critical field to the fully polarized state linearly increases up to the highest $x$ achievable in the present study. Detailed temperature-magnetic field phase diagram study shows that the skyrmion-phase region in the phase diagram becomes larger for large $x$ suggesting larger spin fluctuation; the temperature width of the skyrmion phase region becomes $\Delta T \sim 7$ K, much larger than that in the pure MnSi ($\Delta T \sim 2$ K). Spin-wave stiffness $A$ (or exchange interaction $J$) and Dzyaloshinskii-Moriya interaction $D$ were estimated using magnetization measurement with the aid of mean-field approximation \cite{maleyev,ralph}. The neutron powder diffraction was also performed for selected samples, and shows that the helical modulation vector $k$ has a non-linear behavior. This behavior can be described qualitatively using the mean field approximation. 

\begin{figure*}[t]
\centering
\includegraphics[width=12.12 cm]{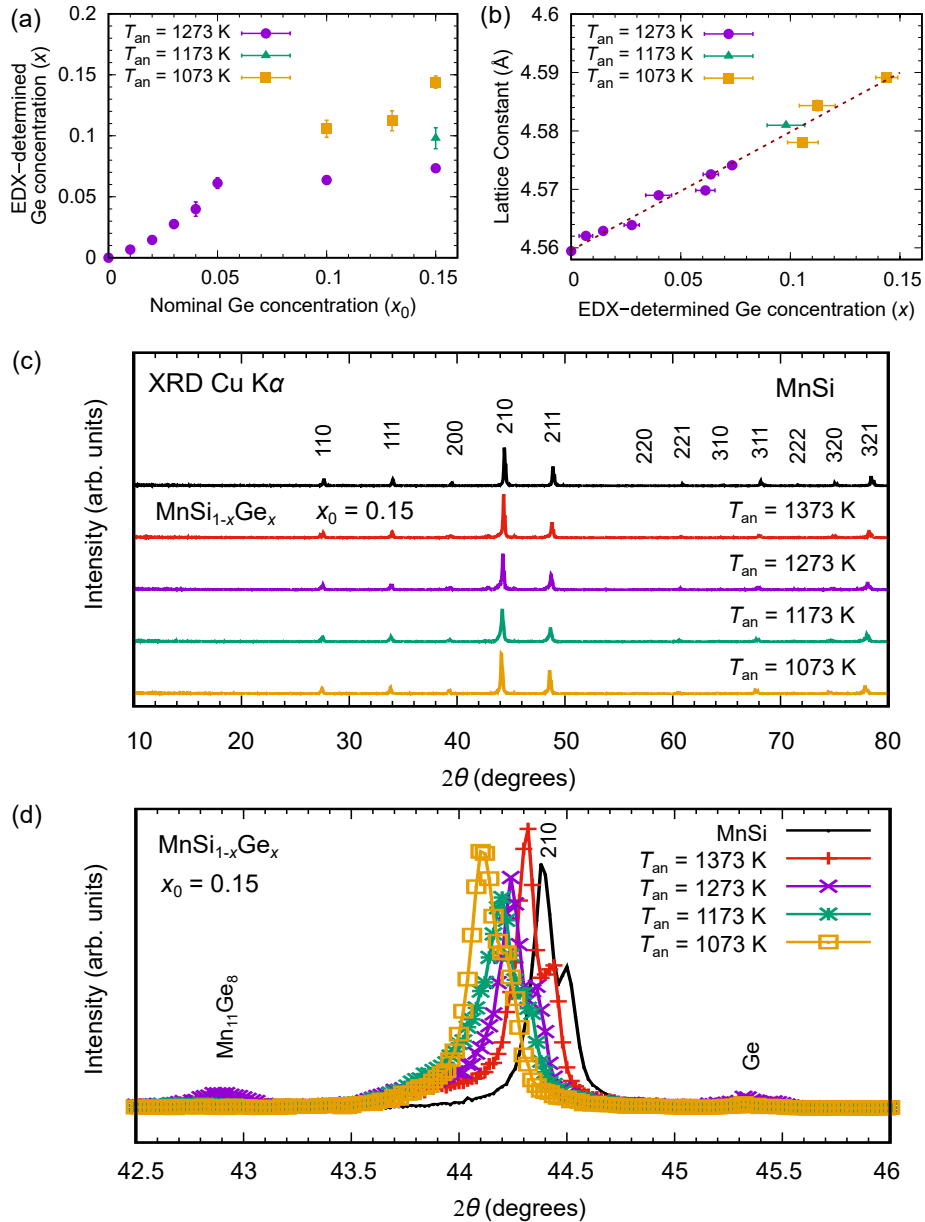}
\caption{(a) The composition determined by the EDX analysis of the main phase of MnSi$_{1-x}$Ge$_x$ alloys annealed at $T_\mathrm{an}=~1073, 1173, \mathrm{and} \: 1273$~K for 5 days, as a function of nominal composition, (b) lattice constants of all prepared samples at different annealing temperatures, as a function of EDX-determined Ge concentration (the dashed line is a guide for the eye), (c) XRD patterns for MnSi (black line) and MnSi$_{1-x}$Ge$_x$ (color lines) at different annealing temperatures $T_\mathrm{an}$, and (d)~Magnified XRD patterns around the 210 reflection.}
\label{fig:allgeanneal}
\end{figure*}

\section{Experimental Details}
The polycrystalline alloy samples of MnSi$_{1-x}$Ge$_{x}$ were prepared by using the arc melting method with high purity elements (Mn: $99.99\%$; Si : $99.999\%$; Ge : $99.999\%$) under titanium-gettered argon atmosphere. The samples were re-melted several times to improve the homogeneity. Some samples with high Ge-substitution (with nominal concentrations $x_0 >$ 0.1) were prepared from arc-melted MnSi and MnGe ingots using an induction furnace to obtain better homogeneity. For the heat treatment or annealing, the samples were put in the Al$_2$O$_3$ crucible, and then sealed in the quartz tube under inert argon gas atmosphere. The annealing was performed using electric furnaces; after ramping to the top temperature (either 1373, 1273, 1173, or 1073 K) with rate $\sim$ 100 K/h, the samples were annealed for 5 days. The samples were then quenched into water. For neutron diffraction experiment, three representative samples were used with initial concentrations $x_0$ = 0.05 and 0.1 annealed at $T$ = 1273 K for 5 days, and $x_0$ = 0.2 annealed at $T$ = 1123 K for 55~hours.    

X-Ray powder diffraction (XRD) was used to identify phases in the obtained polycrystalline samples (Rigaku RINT 2200 with Cu K$\alpha$ radiation, 40 kV $\times$ 30 mA).  XRD confirms that the main phase of obtained polycrystalline sample is the expected B20-type structure. The peak position assignment was performed using the Le-Bail method (Fullprof software, Ref \cite{rodriguez2}); from the obtained peak positions for the $2\theta$ range $10^\circ \leq 2\theta \leq 80^\circ$, the lattice constant $a$ of the Mn(Si,Ge) was estimated using weighted least square method. The microstructure of obtained polycrystalline alloys was checked by taking the backscattered electron (BSE) images using the scanning electron microscope (SEM; Hitachi SU6600).  The incident electron energy was 15 keV.  The elemental compositions of the samples were investigated by performing energy dispersive X-ray (EDX) analysis. 

Magnetization measurements were performed in the temperature range $5 \leq T \leq 300 \: \mathrm{K}$ and in an external magnetic field up to 1 T using a superconducting-quantum interference-device (SQUID) magnetometer (Quantum Design MPMS-XL5). Temperature scans with fixed external field, as well as field scans with fixed temperature were performed depending on the shape of phase boundary in the temperature-field phase diagram.

Neutron diffraction experiment was performed using the cold-neutron triple-axis spectrometer (CTAX) in triple-axis mode, installed at High-Flux Isotope Reactor, Oak Ridge National Laboratory.  To observe low-$q$ (long-wavelength modulation) magnetic peaks, the tight collimations 20'-20' before and after the sample were used. The neutron energy of $E = 3.25$~meV was selected using the PG 002 reflections as monochromator and analyzer. Cooled Be-filter was employed to eliminate the higher harmonic neutrons. The powder samples were sealed in an Al sample can with the $^4$He exchange gas, and the sample can was inserted in the $^4$He vertical-field superconducting cryomagnet.

\section{Experimental Results}
\subsection{Metallurgical survey of solubility limit}

First, we investigate the solubility limit of Ge in MnSi$_{1-x}$Ge$_x$ following Ref. \cite{sivakumar}. The samples with nominal Ge-concentration up to $x_0= 0.15$ were kept in the electrical furnace at annealing temperature $T_{\rm{an}} = 1273$ K for 5 days. The annealed samples were checked by taking BSE images as well as EDX analyses, and the obtained Ge compositions were plotted as a function of the nominal composition in Fig. \ref{fig:allgeanneal}(a) (see filled circle). We found that the solubility limit of Ge at $T_\mathrm{an} = 1273$ K is only around $\sim 0.07$, which is less than that given in Ref. \cite{sivakumar}.

\begin{figure}
\centering
\includegraphics[width = 8.0 cm]{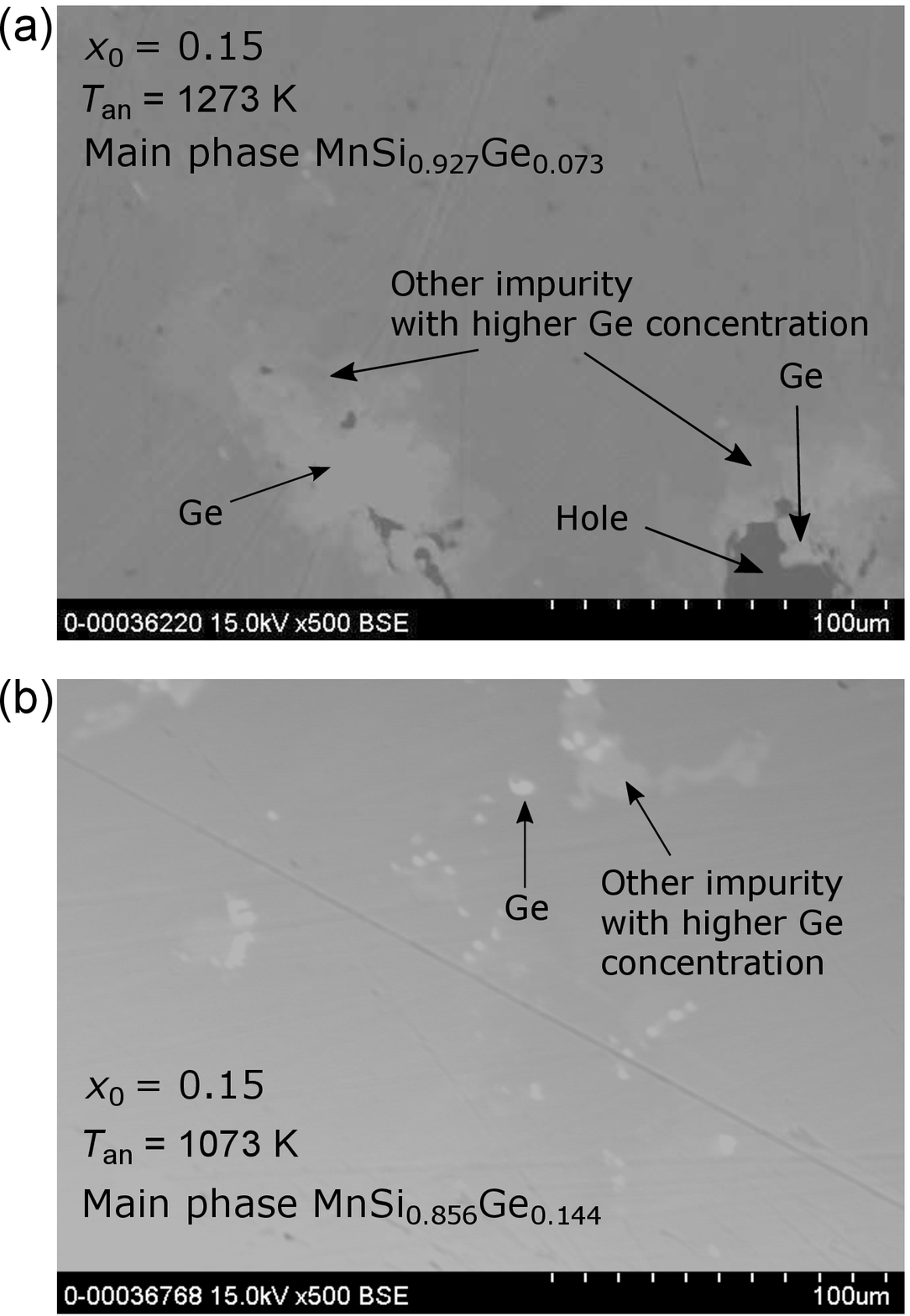}
\caption{BSE images for $x_0= 0.15$ with annealing temperature (a) $T_\mathrm{an}$ = 1273 K, and (b) $T_\mathrm{an}$ = 1073 K.}
\label{fig:bse}
\end{figure}

To check if the higher solubility limit of Ge can be achieved, we performed further metallurgical survey. This metallurgical survey was carried out by fixing the nominal Ge-concentration $x_0 = 0.15$ and changing the annealing temperature $T_{\rm{an}}$ from 1373 to 1073 K. The XRD patterns for the samples with the different annealing temperature are shown in Fig. \ref{fig:allgeanneal}(c). Most of the peaks are indexed with those of the B20 structure. The magnified plot for the selected 2$\theta$ range $42.5^\circ \leq 2\theta \leq 46^\circ$ is also shown in Fig. \ref{fig:allgeanneal}(d). Weak impurity peaks were observed and attributed to the remaining elemental Ge-phase and the other impurity Mn$_{11}$Ge$_8$. It can be seen clearly that the 210 peak appearing at $2\theta \sim 44.3^\circ$ is shifted to the lower angle as the annealing temperature is decreased from 1373 to 1073~K indicating longer lattice constant. Concomitantly, the elemental Ge-phase and Mn$_{11}$Ge$_8$ impurity become weaker. It should be noted that Ge has a greater atomic radius than Si. Hence, the increasing behavior of the lattice constant, as well as the decreasing behavior of the Ge- and Mn$_{11}$Ge$_8$ impurity peaks, suggests higher Ge concentration in the main MnSi$_{1-x}$Ge$_{x}$ phase annealed at the lower temperature. Figures \ref{fig:bse}(a) and \ref{fig:bse}(b) show representative BSE images obtained for annealing temperature $T_{\rm{an}} = 1273$ and 1073~K, respectively. We can see that the sample with $T_{\rm{an}} = 1073$~K has less impurity, indicating that the sample consists of predominant Mn(Si,Ge) phase with small inclusion of impurity Ge. This result, \f{i.e.} the larger solubility range at lower annealing temperature, is somewhat surprising, since the solubility range generally increases at higher temperature due to the entropic effect.  

The Ge-concentrations determined by the EDX analyis of all the low-temperature annealed samples are also shown in Fig. \ref{fig:allgeanneal}(a). The solubility limit of Ge is found to be $x = 0.144(5)$ with annealing temperature $T_{\rm{an}} = 1073$~K. Figure 3 shows the XRD peak profile of the 210 reflection measured on the samples with different nominal Ge concentrations ($x_0$ = 0.15 and 0.2), as well as different annealing time (5 and 20 days). The peak positions are mostly the same for all the preparation conditions. Hence, it can be concluded that the higher nominal concentration ($x_0 = 0.20$) and longer annealing time at $T=1073$~K do not change significantly the solubility limit of Ge. For $x_0 = 0.20$, the Ge concentrations are fluctuating around $\sim 0.15$ with bad homogeneity and large volume fraction of the impurity phase.

The lattice constants of the obtained samples are plotted as a function of the Ge-concentration determined by the EDX analysis in Fig. \ref{fig:allgeanneal}(b). The lattice constant increases linearly as a function of Ge-concentration indicating that the Vegard's law holds for Mn(Si,Ge) system in this composition range. In what follows, we use the EDX-determined Ge concentration $x$, and do not use nominal Ge concentration anymore. 
 
\begin{figure}[t]
\centering
\includegraphics[width = 8.2 cm]{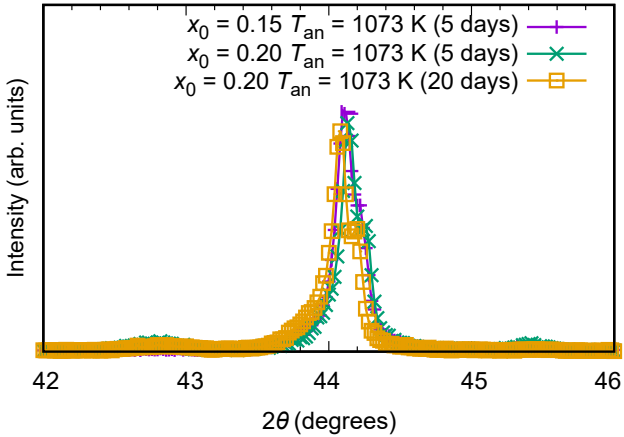}
\caption{Comparison of XRD patterns for different nominal Ge concentration and longer annealing time at $T_\mathrm{an} = 1073$ K.}
\label{fig:xray800}
\end{figure}

\subsection{Ge-concentration dependence of bulk magnetic properties}

%\begin{center}
%\bo{Magnetization measurement}
%\end{center}

\begin{figure}[t]
\includegraphics[width = 7.0 cm]{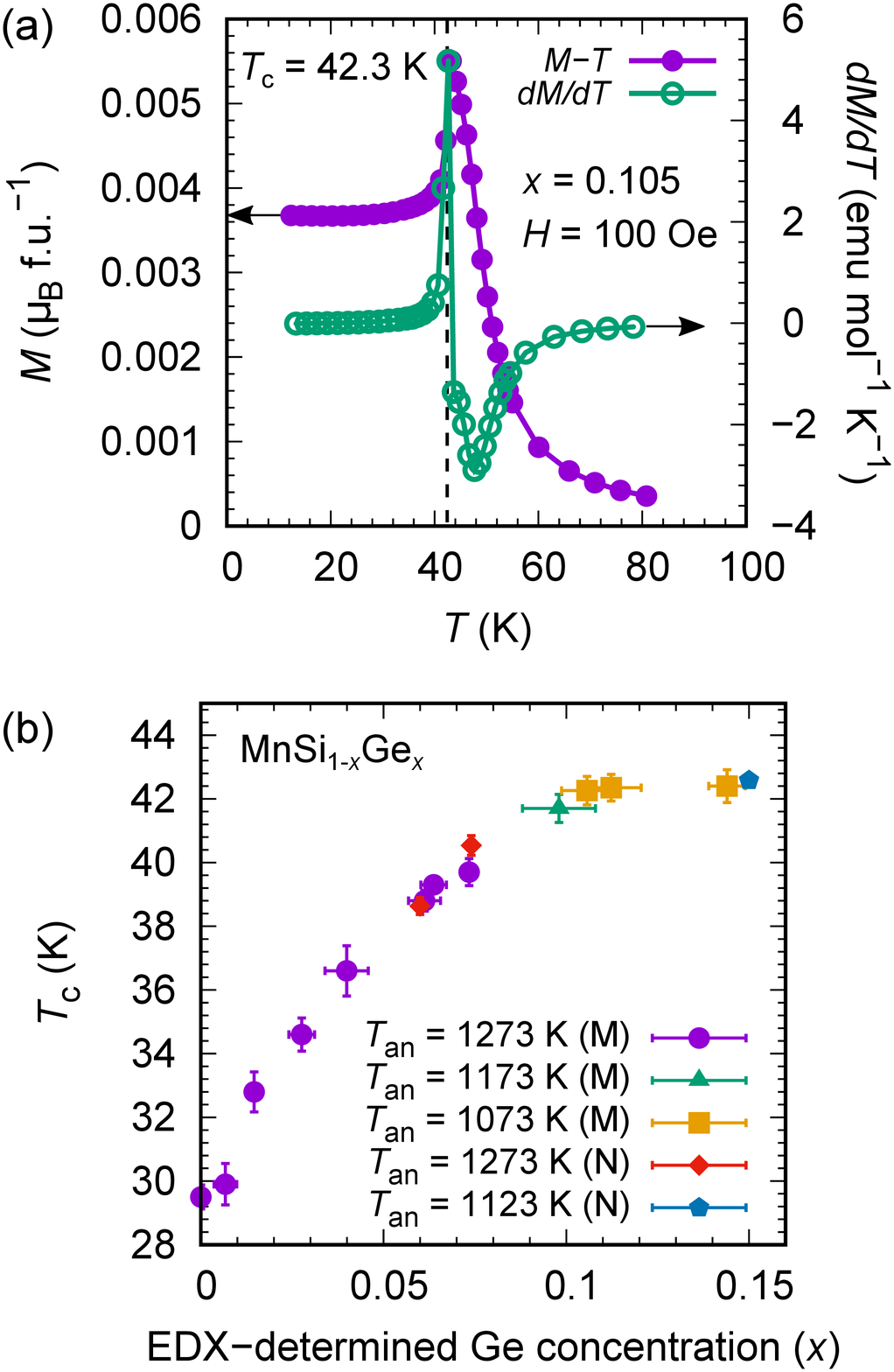}
\caption{(a) $M$-$T$ (filled circles) and $dM/dT$ (open circles) curves under external magnetic field $H = 100$ Oe for $x~= ~0.105(7)$ sample, and (b) Ge-concentration dependence of $T_\mathrm{c}$ obtained from the temperature dependence of magnetization (M) measured under $H = 100$ Oe and neutron experiment (N) under $H = 0$ Oe.}
\label{fig:tc}
\end{figure}

Temperature dependence of the magnetization ($M$-$T$ curve) was measured using the same MnSi$_{1-x}$Ge$_x$ samples prepared and characterized as above.  The magnetization was measured under the external magnetic field $H = 100$~Oe in the temperature range of $10 \leq T \leq 80 \: \mathrm{K}$. The representative result for the $x = 0.105(7)$ sample is shown in Fig. \ref{fig:tc}(a). As can be clearly seen in the figure, the transition temperature derived from $dM/dT$ is $T_\mathrm{c}= 42.3$~K, which is considerably higher than that of the MnSi ($T_\mathrm{c} = 29.5$~K). The Ge-concentration dependence of $T_\mathrm{c}$, determined similarly for all the samples with different concentrations, is plotted in Fig. \ref{fig:tc} (b). It is shown that the helical ordering temperature $T_\mathrm{c}$ increases rapidly at the low concentration $x<0.1$ and becomes saturated at the high concentration range $x>0.1$. 

Next, we investigated magnetic field dependence of the magnetization ($M$-$H$ curve) at fixed temperatures. For this measurement, we need to carefully consider the sample shape since the magnetic property such as critical field is sensitive to the sample shape due to the demagnetizing field effect \cite{bauer}. Shape of the samples used in the magnetization measurement can be reasonably assumed as rectangular parallelepiped with a small size fluctuation from sample to sample.  The size distribution for the width ($w$), length ($l$) and thickness ($t$) is $1.8\leq w \leq 3.1$~mm (average 2.5~mm), $3.2\leq l \leq 4.8 $~mm (average 3.75~mm), and $0.4\leq t \leq 0.8 $~mm (average 0.6~mm), respectively. We set the shorter axis (thickness direction) perpendicular to the magnetic field to minimize the demagnetization effect.
The demagnetizing factor $D_{\rm ext}$ for the rectangular parallelepiped samples under the external field can be estimated numerically following Ref. \cite{aharoni}.
The correction factor which relates the applied external field to true internal magnetic field as $H_{\rm int} = f H_{\rm ext}$ can be obtained as $f=(1- D_{\rm ext} \chi_\mathrm{con}^\mathrm{ext})$, where $\chi_\mathrm{con}^\mathrm{ext}$ stands for the conical susceptibility defined as $M = \chi_\mathrm{con}^\mathrm{ext} H_\mathrm{ext}$ \cite{Bauer2016book}, estimated at $H_{\rm ext}  = 3.5\;\mathrm{kOe}$. The smallest correction factor was obtained as $f_{\rm min} =0.953$ for the sample with $w = 2.9$, $l = 3.7$, and $t = 0.8$~mm, whereas the largest one as $f_{\rm max} =0.975$ for $w = 3.1$, $l = 4.8$, and $t = 0.4$~mm.
Since even for the two extreme cases the variation is sufficiently small as only 2.2~\%, we ignore the demagnetization field effect in the following, and show the uncorrected data.

\begin{figure}[t]
\includegraphics[width = 7.4 cm]{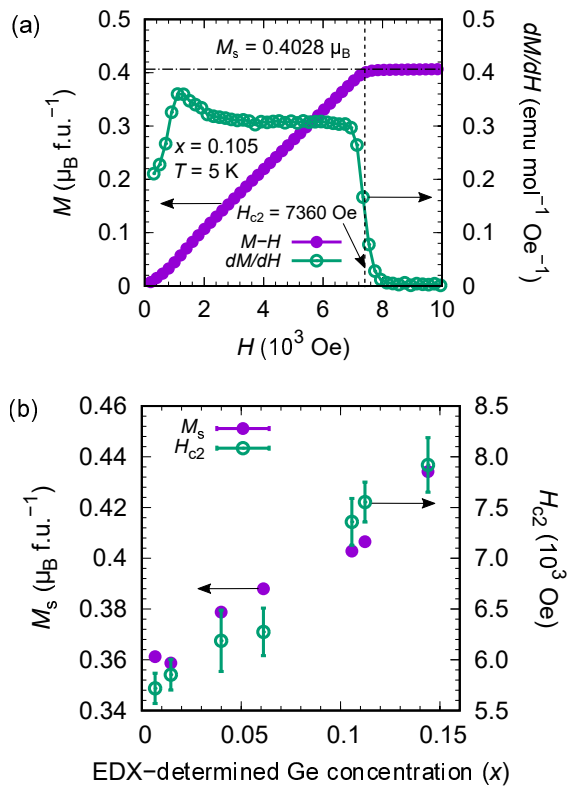}
\caption{(a) $M$-$H$ (filled circles) and $dM/dH$ (open circles) curves at $T = 5$ K for the $x = 0.105(7)$ sample, and (b) Ge-concentration dependence of saturation moment $M_\mathrm{s}$ and critical field $H_\mathrm{c2}$.}
\label{fig:mh0p1}
\end{figure}

\begin{figure*}[t]
\includegraphics[width = 14cm]{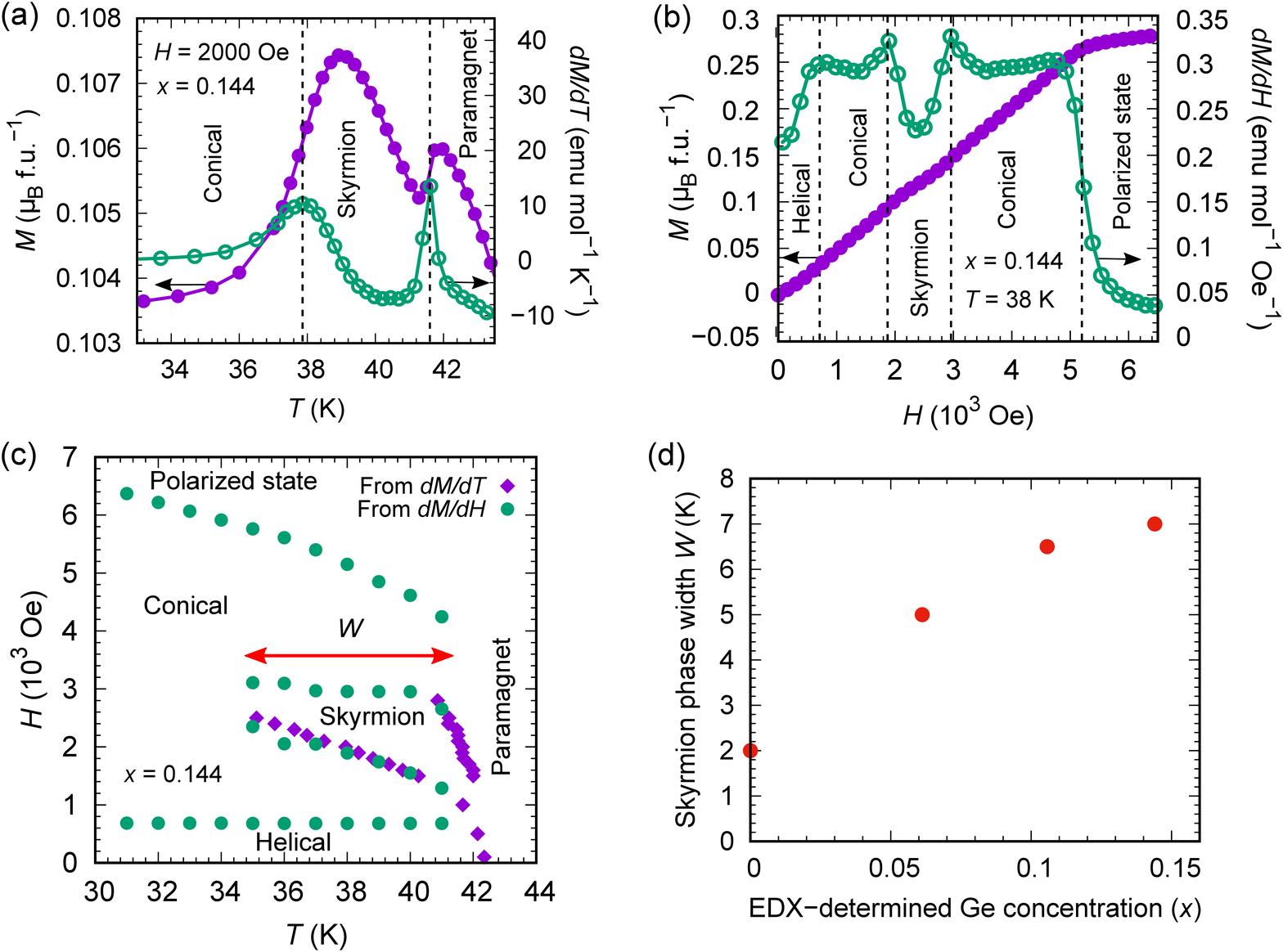}
\caption{(a) $M$-$T$ (filled circles) and $dM/dT$ (open circles) curves with external magnetic field $H=2000$ Oe, (b) $M$-$H$ (filled circles) and $dM/dH$ (open circles) curves with temperature $T = 38$ K, (c) full phase diagram for representing sample $x=0.144(5)$, and (d) Ge-concentration dependence of the width of the skyrmion-phase region $W$.}
\label{fig:phase0p15}
\end{figure*}
Representative $M$-$H$ and $dM/dH$ curves obtained at $T=5$ K for $x=0.105(7)$ are shown in Fig. \ref{fig:mh0p1}(a). The $M$-$H$ curve shows monotonic increase for increasing $H$ with clear saturation at high field. By linearly extrapolating the $H$-dependence of the magnetization in the higher $H$ range ($H> 8000$ Oe) to $H\rightarrow0$, we estimated the size of the saturation moment $M_s$ for the sample as 0.4028(1)~$\mu_\mathrm{B}$ [the value in the parentheses is an uncertainty range (one sigma) obtained in the fitting]. From $dM/dH$ curve, we estimate the critical field to the fully polarized state $H_\mathrm{c2}$ as $7360 \: \mathrm{Oe}$. The saturation moment $M_\mathrm{s}$ and critical field $H_\mathrm{c2}$ for all the samples are estimated in similar manner and plotted in Fig. \ref{fig:mh0p1}(b). The saturation moment per Mn ion shows linear increase as Ge-concentrations increases. The increase of saturation moment $M_\mathrm{s}$ is a typical behavior for the effect of negative chemical pressure as it also can be found in Al- or Ga-substituted MnSi \cite{dhital1}. However, Al- or Ga-substitution tends to decrease the critical field to the fully polarized state $H_\mathrm{c2}$ in stark contrast to Ge-substitution. As shown in Fig. \ref{fig:mh0p1}(b), Ge-substitution increases the critical field $H_\mathrm{c2}$.

\begin{figure*}[t]
\centering
\includegraphics[width = 14cm]{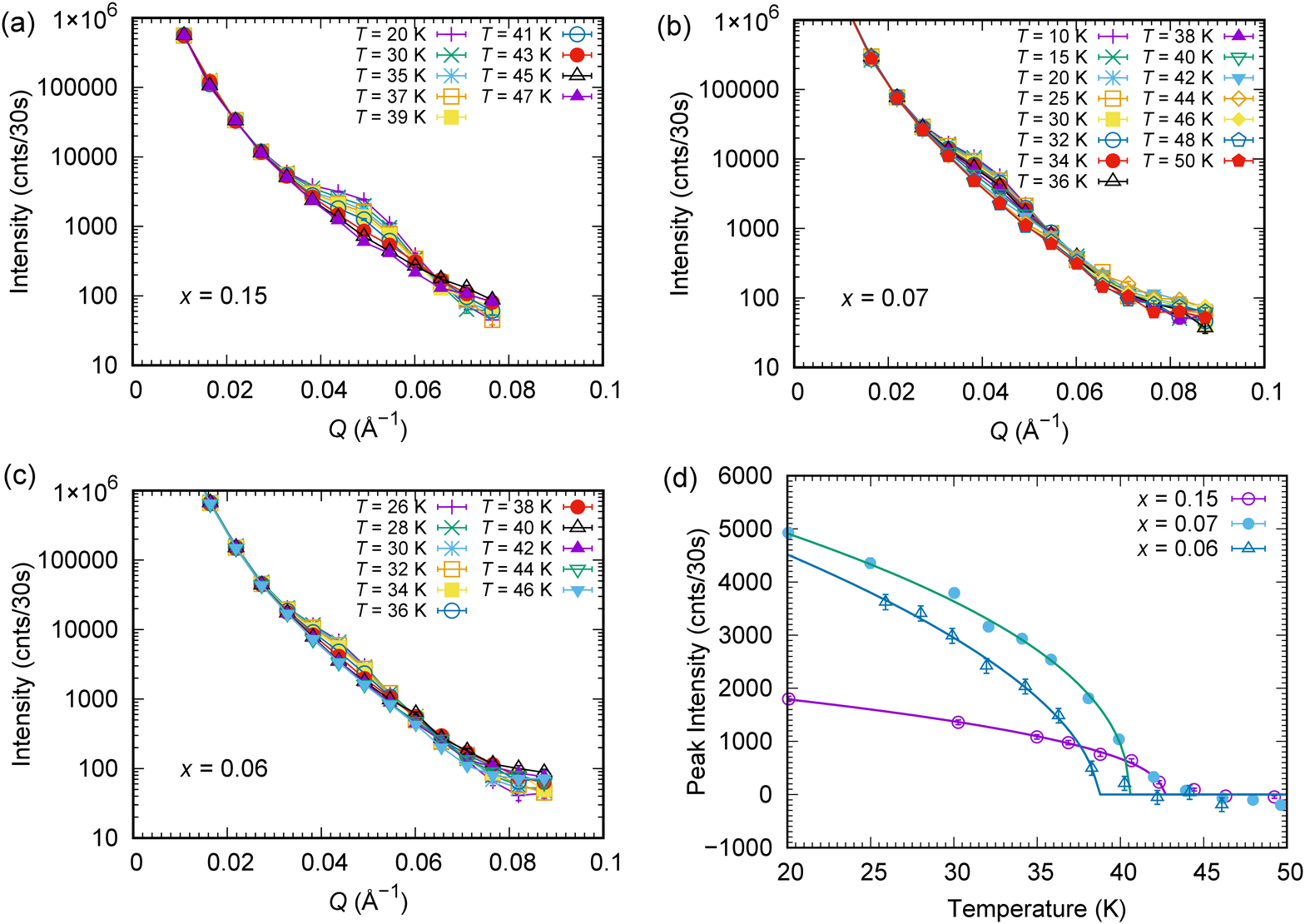}
\caption{(a-c) Neutron diffraction patterns in the small $Q$ range for the three MnSi$_{1-x}$Ge$_x$ samples with (a) $x$ = 0.15, (b) 0.07, and (c) 0.06.  The temperatures were given in the figures. (d) Temperature dependence of the peak intensity for (open circles) $x$ = 0.15 , (filled circles) 0.07, and (open triangles) 0.06.  The transition temperatures were estimated as $T_\mathrm{c}$ = 42.6(1), 40.5(3), and 38.6(3) K for $x$ = 0.15, 0.07, and 0.06, respectively.  The background, estimated at high temperature paramagnetic data, is subtracted from the data.  The solid lines are the results of the fitting to the power law.}
\label{fig:sato1-4}
\end{figure*}

Temperature-field phase diagrams for selected samples were investigated using the various $M$-$T$ and $M$-$H$ scans. Representative $M$-$T$ and $dM/dT$ results for $x = 0.144(5)$ are shown in Fig. \ref{fig:phase0p15}(a) with constant magnetic field $H=2000$ Oe, whereas those for $M$-$H$ and $dM/dH$ in Fig. \ref{fig:phase0p15}(b) with constant temperature $T = 38$ K. The transitions between the paramagnetic, fully-polarized, helical, conical, and skyrmion phases are indicated by dashed lines. The phase transitions are summarized in the $H$-$T$ phase diagram shown in Fig. \ref{fig:phase0p15}(c). From the phase diagram, one can see that the effect of the Ge-substitution is not only the change of the helical ordering temperature $T_\mathrm{c}$ but also enlargement of the skyrmion phase region. Typically, for MnSi, the temperature width of the skyrmion-phase region is about $2\sim3$ K as reported in earlier works \cite{muhlbauer,nadya}. In contrast, the width of the skyrmion-phase region increases in Ge-substituted MnSi, and is about 7 K for $x = 0.144(5)$. The widths of skyrmion phase for the four representative samples with the estimated Ge-concentrations $x$ = 0, 0.061(4), 0.105(7), and 0.144(5), are plotted in Fig. \ref{fig:phase0p15}(d). The skyrmion width seemingly behaves like critical temperature $T_\mathrm{c}$ showing the saturation at high concentration.

\begin{figure}[t]
\centering
\includegraphics[width = 8.2cm]{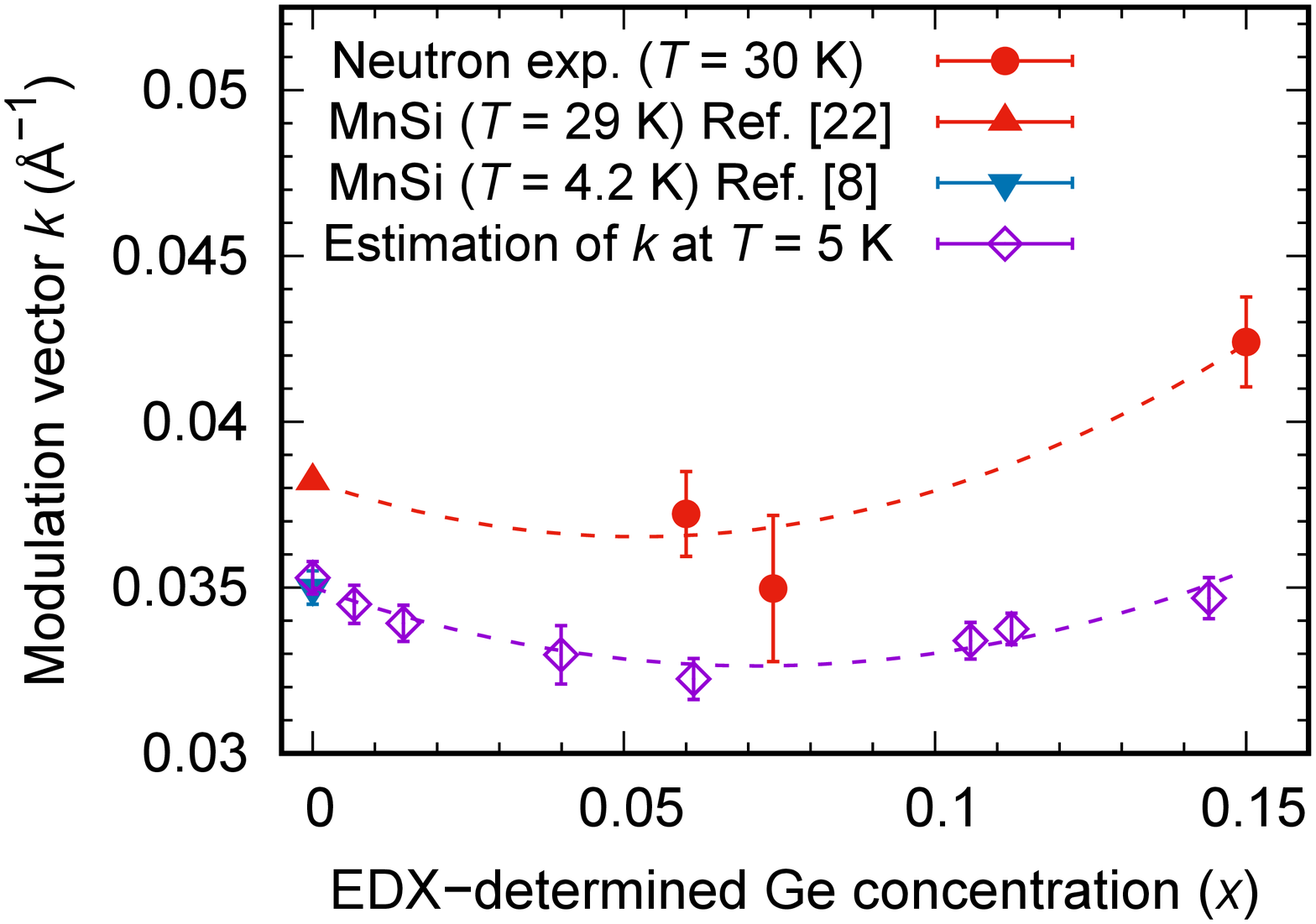}
\caption{Ge concentration dependence of the modulation vectors $k$ estimated from the neutron diffraction data at $T=30$~K shown in Fig. \ref{fig:sato1-4} (filled circles) for three representative MnSi$_{1-x}$Ge$_x$ samples. The modulation vectors of MnSi are obtained at $T = 29$~K (Ref. \cite{gregorievmnsi}, filled triangle up) and $T=4.2$~K (Ref. \cite{ishikawa}, filled triangle down). The calculated modulation vectors $k$ derived from magnetization measurement at $T = 5$~K are shown by open diamond as described in the discussion. The dashed lines are a guide for the eye.}
\label{fig:mod}
\end{figure}

\subsection{Neutron diffraction}
Finally, the neutron diffraction experiments were performed for the three representative samples with the estimated Ge-concentrations $x$ = 0.06, 0.07, and 0.15. The low-$Q$ diffraction patterns for the three samples are shown in Fig. \ref{fig:sato1-4}. There appears significantly increasing background for $Q \rightarrow 0$ due to the contamination from the direct beam. Nevertheless, it is clearly seen that in all the samples, the magnetic Bragg peak develops at low temperatures. By fitting each patterns to the Gaussian function together with the background slope estimated from the high-temperature paramagnetic data, the intensity and position of the helical Bragg peak were estimated as a function of the temperature. The temperature dependence of the peak intensity is shown in Fig. \ref{fig:sato1-4}(d). The constant temperature background, estimated from the high-temperature paramagnetic contribution, was removed. The transition temperature was estimated for each sample, by fitting to the empirical power law function $I(T) \propto I_0 (T_\mathrm{c} - T)^{2\beta}$, and is also shown in Fig. \ref{fig:tc}(b). The transition temperature decreases for $x\rightarrow 0$, in good agreement with the macroscopic measurement. Shown in the Fig. \ref{fig:mod} is Ge concentration dependence of the resulting modulation vector estimated from the peak position at $T = 30$ K. Although the uncertainly is quite large due to the large contaminating background, it may be seen that the modulation vector stays almost unchanged up to $x=0.08$ and then starts to increase for larger $x$.

\section{Discussion}

In the present study, we have performed extensive metallurgical survey and found that the Si can be substituted by Ge up to $x=0.144(5)$ using ambient pressure synthesis technique. The resulting high quality polycrystalline samples were used in the detailed bulk magnetic measurement, as well as neutron diffraction. In the present work, the clear Ge-concentration dependence of helical ordering temperature $T_\mathrm{c}$ is obtained up to $x\sim0.15$. $T_\mathrm{c}$ increases rapidly at small $x$, and becomes saturated at the high concentration range $x>0.1$. The Ge-substitution also increases both the saturation moment per Mn ion ($M_\mathrm{s}$) and the critical field to the polarized state ($H_\mathrm{c2}$), but linearly. The increase of $M_\mathrm{s}$ is a typical behavior for the effect of negative chemical pressure as it can be also found in Al- or Ga-substituted MnSi \cite{dhital1}. However, Al-or Ga-substitution tends to decrease $H_\mathrm{c2}$.

First, we discuss the implication of these bulk parameters using the well established model. A simplified model Hamiltonian for MnSi may be written as  \cite{maleyev,chizhikov}:

\begin{equation}
H= \frac{1}{2} \sum_{<ij>} \left(-J\bo{S}_i \cdot \bo{S}_j +  \bo{D}_{ij}\cdot (\bo{S}_i \times \bo{S}_j) \right) + \sum_{i} \bo{h} \cdot \bo{S}_i \: .
\end{equation}

\noindent Here, the total spin of the unit cell is denoted by $S$, and is used as a magnetic entity. Note that since there are four magnetic ions per unit cell, $S$ is related to the cell saturation magnetization $M_\mathrm{s}=g\mu_B S$, where $g\simeq 2$ and $\mu_B>0$ are the $g$ factor and Bohr magneton, respectively. For example, the saturation magnetic moment per spin for $x=0.144(5)$ sample is $0.434$ $\mu_B$, and hence the unit-cell spin is $S=0.217\times 4 = 0.87$. The sum over $i$ is taken over all the total spins of unit cells, while the sum over $j$ is taken over six neighboring unit cell spins. The first, second, and the last term stand for the exchange interaction (exchange parameter: $J$), Dzyaloshinskii-Moriya (DM) interaction (DM parameter: $D$, DM vector is assumed to be parallel to the helical propagation vector), and Zeeman energy ($\bo{h} = g\mu_B\bo{H}_\mathrm{ext}$, where $\bo{H}_\mathrm{ext}$ is the external field).
By using the classical ground-state energy calculation as described in Ref. \cite{maleyev}, the helical ordering modulation vector $k$ and the critical field to the fully polarized state $H_\mathrm{c2}$ are given as:
\begin{equation}
k=D/aJ
\label{eq:2}
\end{equation}
\begin{equation}
H_\mathrm{c2} = \frac{Ak^2}{g\mu_\mathrm{B}}
\label{eq:3}
\end{equation}

\noindent where $A = JSa^2$ is so called the spin-wave stiffness parameter and $a$ is the lattice constant.

In addition, in the mean field approximation by considering the Heisenberg-exchange and DM interaction, the critical temperature $T_\mathrm{c}$ is written as \cite{ralph}: 
\begin{equation}
T_\mathrm{c} = \frac{JzS(S+1)}{3k_\mathrm{B}}\left(1+\frac{1}{z}\frac{D^2}{J^2}\right)\:,
\label{eq:4}
\end{equation}

\noindent since $D \ll J$ for MnSi and the second term is already $O \left( (\frac{D}{J})^2 \right)$, then we ignored the higher order term. Thus, the equation (\ref{eq:4}) can be simplified in terms of spin-wave stiffness A as follows: 
\begin{equation}
T_\mathrm{c} = \gamma \frac{A}{k_Ba^2}\: .
\label{eq:5}
\end{equation}

\begin{figure}[b]
\centering
\includegraphics[width = 6.5cm]{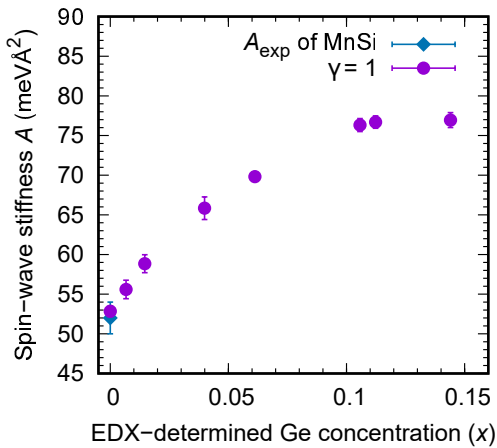}
\caption{Ge concentration dependence of the spin-wave stiffness $A$. The observed spin-wave stiffness $A_\mathrm{exp}$ of MnSi is obtained from Ref. \cite{ishikawa2,grigoriev02}}
\label{fig:A1A2}
\end{figure}

\noindent In the mean field approximation, $\gamma = z(S+1)/3$, with z being the coordination number (for cubic crystal $z = 6$). It is, however, well known that the mean-field theory considerably overestimates $T_\mathrm{c}$ by ignoring spin fluctuations, which are known to be significant in MnSi \cite{papas}. Indeed, in the earlier work, it was found that $T_\mathrm{c}$ is well scaled by $A$ with $\gamma \sim 1$ for Mn$_{1-x}$Fe$_x$Si \cite{grigoriev01}.  Therefore, we also assume $\gamma \sim 1$ for MnSi$_{1-x}$Ge$_x$ in the following.
 
The estimation of spin-wave stiffness as a function of Ge-concentration is shown in Fig. \ref{fig:A1A2}. The spin-wave stiffness $A$ increases in the low $x$ range, while it becomes saturated at high Ge-concentration. This behavior is in stark contrast to that in Mn$_{1-x}$Fe$_{x}$Si where $A$ decreases monotonically \cite{grigoriev01}. To confirm the increasing behavior of spin-wave stiffness $A$ of Ge-doped MnSi in particular, the saturation at $x>0.1$, inelastic neutron experiment for MnSi$_{1-x}$Ge$_x$ is necessary in the future. 

\begin{figure}[t]
\includegraphics[width = 7.5cm]{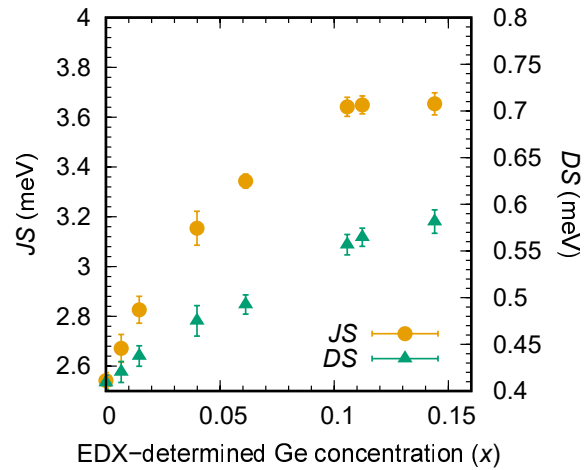}
\caption{Ge-concentration dependence of exchange ($JS$) and Dzyaloshinskii-Moriya ($DS$) parameters.}
\label{fig:fundpara}
\end{figure}

From Eq. \ref{eq:2} and \ref{eq:3}, as well as $A$ obtained above, we estimate the exchange interaction $JS$, DM interaction $DS$, and the helical modulation vector $k$. The Ge-concentration dependence of the exchange and DM interactions are plotted in Fig.~\ref{fig:fundpara}. $JS$ shows saturation behavior, whereas $DS$ increases linearly, reflecting the $x$-dependence of $T_\mathrm{c}$, and $M_\mathrm{s}$ or $H_\mathrm{c2}$, respectively. The estimated helical modulation vector $k$ as a function of the Ge-concentration is plotted in Fig.~\ref{fig:mod}. First, we note that the estimated $k = 0.0352(4)$ \r{A}$^{-1}$ at $T = 5$ K in the above is in good agreement with the reported $k$ obtained in the neutron diffraction at $T = 4.2$ K \cite{ishikawa}. This suggests the validity of the modulation vector estimated in the mean field approximation. 
In the presence of Ge-substitution, the estimated modulation vector decreases in the beginning then starts to increase at higher $x$. Our neutron measurement was done at $T=30$ K due to technical issue and hence we cannot directly compare the neutron results to the estimated $k$. Nonetheless, one may note that the $x$-dependence of the neutron observed $k$ is qualitatively consistent with the estimation; it shows slightly decreasing behavior in the low $x$ range, and then starts to increase at higher $x$. 

It is, however, noted that the increase of $k$ at higher $x$ may be weaker in the mean field estimation than the neutron observation. This indicates that there should be an intrinsic-enhancement of spin fluctuations which are in principle ignored in the mean field approximation. It should be noted that for large $x$ the width of skyrmion-phase region becomes larger as shown in Fig. \ref{fig:phase0p15}(d). Since it is the fluctuation effect beyond the mean field level that stabilizes skyrmion phase, the large spin fluctuation is also suggested from this observation. Another possibility may be the enhanced long range RKKY interaction as inferred in the Ref. \cite{fujishiro}.
Since this interaction can control $k$ vector independently from the DM interaction, this can certainly explain the stronger increase of $k$ at large $x$ if properly introduced. From the present experimental data we cannot conclude which (or both) is the origin of the increasing behavior of the modulation vector $k$. Further study, in particular, theoretical calculation, is necessary to elucidate this issue.

\section{Conclusions}
MnSi$_{1-x}$Ge$_x$ polycrystalline samples were successfully synthesized. The solubility limit of Ge under ambient pressure is found as $x=0.144(5)$ with the annealing temperature $T_\mathrm{an} = 1073$ K. The lattice constant linearly increases with the Ge-substitution in accordance with Vegard's law. We extracted various magnetic parameters including $M_\mathrm{s}$, $T_\mathrm{c}$, and $H_\mathrm{c2}$ from the bulk magnetic measurements and are able to estimate the helical modulation vector $k$ at $T = 5$ K. The estimated helical modulation vector decreases in the low $x$-range, and then starts to increase at higher $x$, which is qualitatively consistent with the observed $k$ at $T = 30$ K in the present neutron diffraction. The width of the skyrmion phase region shows increasing behavior for large $x$, suggesting increased spin fluctuation as $x$ becomes large.

\begin{center}
\bo{ACKNOWLEDGMENTS}
\end{center}

The authors thank S. Ohhashi and M. Ozawa for technical assistance in using Electron Scanning Microscope (SEM; Hitachi SU6600).  This work was partly supported by Grants-In-Aid for Scientific Research (15H05883, 17K18744, 19H01834, 19K21839 and JP19H05824). A portion of this research used resources at the High Flux Isotope Reactor, a DOE Office of Science User Facility operated by the Oak Ridge National Laboratory, 
%The work at HFIR, Oak Ridge National Laboratory was sponsored by the Division of Scientific User Facilities, Office of Basic Energy Science, U.S. Department of Energy (DOE), 
and was partly supported by the US-Japan Collaborative Program on Neutron Scattering.  The work at IMRAM was partly supported by by the research program “Dynamic alliance for open innovation bridging human, environment, and materials”.

\bibliographystyle{apsrev4-1}
%\bibliographystyle{aipnum4-1}
%\bibliography{reference}

\begin{thebibliography}{26}%
\makeatletter
\providecommand \@ifxundefined [1]{%
 \@ifx{#1\undefined}
}%
\providecommand \@ifnum [1]{%
 \ifnum #1\expandafter \@firstoftwo
 \else \expandafter \@secondoftwo
 \fi
}%
\providecommand \@ifx [1]{%
 \ifx #1\expandafter \@firstoftwo
 \else \expandafter \@secondoftwo
 \fi
}%
\providecommand \natexlab [1]{#1}%
\providecommand \enquote  [1]{``#1''}%
\providecommand \bibnamefont  [1]{#1}%
\providecommand \bibfnamefont [1]{#1}%
\providecommand \citenamefont [1]{#1}%
\providecommand \href@noop [0]{\@secondoftwo}%
\providecommand \href [0]{\begingroup \@sanitize@url \@href}%
\providecommand \@href[1]{\@@startlink{#1}\@@href}%
\providecommand \@@href[1]{\endgroup#1\@@endlink}%
\providecommand \@sanitize@url [0]{\catcode `\\12\catcode `\$12\catcode
  `\&12\catcode `\#12\catcode `\^12\catcode `\_12\catcode `\%12\relax}%
\providecommand \@@startlink[1]{}%
\providecommand \@@endlink[0]{}%
\providecommand \url  [0]{\begingroup\@sanitize@url \@url }%
\providecommand \@url [1]{\endgroup\@href {#1}{\urlprefix }}%
\providecommand \urlprefix  [0]{URL }%
\providecommand \Eprint [0]{\href }%
\providecommand \doibase [0]{http://dx.doi.org/}%
\providecommand \selectlanguage [0]{\@gobble}%
\providecommand \bibinfo  [0]{\@secondoftwo}%
\providecommand \bibfield  [0]{\@secondoftwo}%
\providecommand \translation [1]{[#1]}%
\providecommand \BibitemOpen [0]{}%
\providecommand \bibitemStop [0]{}%
\providecommand \bibitemNoStop [0]{.\EOS\space}%
\providecommand \EOS [0]{\spacefactor3000\relax}%
\providecommand \BibitemShut  [1]{\csname bibitem#1\endcsname}%
\let\auto@bib@innerbib\@empty
%</preamble>
\bibitem [{\citenamefont {Jeong}\ and\ \citenamefont {Pickett}(2004)}]{jeong}%
  \BibitemOpen
  \bibfield  {author} {\bibinfo {author} {\bibfnamefont {T.}~\bibnamefont
  {Jeong}}\ and\ \bibinfo {author} {\bibfnamefont {W.~E.}\ \bibnamefont
  {Pickett}},\ }\href {\doibase 10.1103/PhysRevB.70.075114} {\bibfield
  {journal} {\bibinfo  {journal} {Phys. Rev. B}\ }\textbf {\bibinfo {volume}
  {70}},\ \bibinfo {pages} {075114} (\bibinfo {year} {2004})}\BibitemShut
  {NoStop}%
\bibitem [{\citenamefont {Thessieu}\ \emph {et~al.}(1995)\citenamefont
  {Thessieu}, \citenamefont {Flouquet}, \citenamefont {Lapertot}, \citenamefont
  {Stepanov},\ and\ \citenamefont {Jaccard}}]{thessieu}%
  \BibitemOpen
  \bibfield  {author} {\bibinfo {author} {\bibfnamefont {C.}~\bibnamefont
  {Thessieu}}, \bibinfo {author} {\bibfnamefont {J.}~\bibnamefont {Flouquet}},
  \bibinfo {author} {\bibfnamefont {G.}~\bibnamefont {Lapertot}}, \bibinfo
  {author} {\bibfnamefont {A.}~\bibnamefont {Stepanov}}, \ and\ \bibinfo
  {author} {\bibfnamefont {D.}~\bibnamefont {Jaccard}},\ }\href {\doibase
  https://doi.org/10.1016/0038-1098(95)00356-8} {\bibfield  {journal} {\bibinfo
   {journal} {Solid State Commun.}\ }\textbf {\bibinfo {volume} {95}},\
  \bibinfo {pages} {707 } (\bibinfo {year} {1995})}\BibitemShut {NoStop}%
\bibitem [{\citenamefont {Nakajima}\ \emph {et~al.}(2017)\citenamefont
  {Nakajima}, \citenamefont {Oike}, \citenamefont {Kikkawa}, \citenamefont
  {Gilbert}, \citenamefont {Booth}, \citenamefont {Kakurai}, \citenamefont
  {Taguchi}, \citenamefont {Tokura}, \citenamefont {Kagawa},\ and\
  \citenamefont {Arima}}]{nakajima}%
  \BibitemOpen
  \bibfield  {author} {\bibinfo {author} {\bibfnamefont {T.}~\bibnamefont
  {Nakajima}}, \bibinfo {author} {\bibfnamefont {H.}~\bibnamefont {Oike}},
  \bibinfo {author} {\bibfnamefont {A.}~\bibnamefont {Kikkawa}}, \bibinfo
  {author} {\bibfnamefont {E.~P.}\ \bibnamefont {Gilbert}}, \bibinfo {author}
  {\bibfnamefont {N.}~\bibnamefont {Booth}}, \bibinfo {author} {\bibfnamefont
  {K.}~\bibnamefont {Kakurai}}, \bibinfo {author} {\bibfnamefont
  {Y.}~\bibnamefont {Taguchi}}, \bibinfo {author} {\bibfnamefont
  {Y.}~\bibnamefont {Tokura}}, \bibinfo {author} {\bibfnamefont
  {F.}~\bibnamefont {Kagawa}}, \ and\ \bibinfo {author} {\bibfnamefont {T.-h.}\
  \bibnamefont {Arima}},\ }\href
  {https://advances.sciencemag.org/content/3/6/e1602562} {\bibfield  {journal}
  {\bibinfo  {journal} {Sci. Adv.}\ }\textbf {\bibinfo {volume} {3}} (\bibinfo
  {year} {2017})}\BibitemShut {NoStop}%
\bibitem [{\citenamefont {Tite}\ \emph {et~al.}(2010)\citenamefont {Tite},
  \citenamefont {Shu}, \citenamefont {Chou},\ and\ \citenamefont
  {Chang}}]{tite}%
  \BibitemOpen
  \bibfield  {author} {\bibinfo {author} {\bibfnamefont {T.}~\bibnamefont
  {Tite}}, \bibinfo {author} {\bibfnamefont {G.~J.}\ \bibnamefont {Shu}},
  \bibinfo {author} {\bibfnamefont {F.~C.}\ \bibnamefont {Chou}}, \ and\
  \bibinfo {author} {\bibfnamefont {Y.-M.}\ \bibnamefont {Chang}},\ }\href
  {https://doi.org/10.1063/1.3464980} {\bibfield  {journal} {\bibinfo
  {journal} {Appl. Phys. Lett.}\ }\textbf {\bibinfo {volume} {97}},\ \bibinfo
  {pages} {031909} (\bibinfo {year} {2010})}\BibitemShut {NoStop}%
\bibitem [{\citenamefont {Neubauer}\ \emph {et~al.}(2009)\citenamefont
  {Neubauer}, \citenamefont {Pfleiderer}, \citenamefont {Binz}, \citenamefont
  {Rosch}, \citenamefont {Ritz}, \citenamefont {Niklowitz},\ and\ \citenamefont
  {B\"oni}}]{neubauer}%
  \BibitemOpen
  \bibfield  {author} {\bibinfo {author} {\bibfnamefont {A.}~\bibnamefont
  {Neubauer}}, \bibinfo {author} {\bibfnamefont {C.}~\bibnamefont
  {Pfleiderer}}, \bibinfo {author} {\bibfnamefont {B.}~\bibnamefont {Binz}},
  \bibinfo {author} {\bibfnamefont {A.}~\bibnamefont {Rosch}}, \bibinfo
  {author} {\bibfnamefont {R.}~\bibnamefont {Ritz}}, \bibinfo {author}
  {\bibfnamefont {P.~G.}\ \bibnamefont {Niklowitz}}, \ and\ \bibinfo {author}
  {\bibfnamefont {P.}~\bibnamefont {B\"oni}},\ }\href {\doibase
  10.1103/PhysRevLett.102.186602} {\bibfield  {journal} {\bibinfo  {journal}
  {Phys. Rev. Lett.}\ }\textbf {\bibinfo {volume} {102}},\ \bibinfo {pages}
  {186602} (\bibinfo {year} {2009})}\BibitemShut {NoStop}%
\bibitem [{\citenamefont {Bauer}\ and\ \citenamefont
  {Pfleiderer}(2012)}]{bauer}%
  \BibitemOpen
  \bibfield  {author} {\bibinfo {author} {\bibfnamefont {A.}~\bibnamefont
  {Bauer}}\ and\ \bibinfo {author} {\bibfnamefont {C.}~\bibnamefont
  {Pfleiderer}},\ }\href {\doibase 10.1103/PhysRevB.85.214418} {\bibfield
  {journal} {\bibinfo  {journal} {Phys. Rev. B}\ }\textbf {\bibinfo {volume}
  {85}},\ \bibinfo {pages} {214418} (\bibinfo {year} {2012})}\BibitemShut
  {NoStop}%
\bibitem [{\citenamefont {Williams}\ \emph {et~al.}(1966)\citenamefont
  {Williams}, \citenamefont {Wernick}, \citenamefont {Sherwood},\ and\
  \citenamefont {Wertheim}}]{williams}%
  \BibitemOpen
  \bibfield  {author} {\bibinfo {author} {\bibfnamefont {H.~J.}\ \bibnamefont
  {Williams}}, \bibinfo {author} {\bibfnamefont {J.~H.}\ \bibnamefont
  {Wernick}}, \bibinfo {author} {\bibfnamefont {R.~C.}\ \bibnamefont
  {Sherwood}}, \ and\ \bibinfo {author} {\bibfnamefont {G.~K.}\ \bibnamefont
  {Wertheim}},\ }\href {\doibase 10.1063/1.1708422} {\bibfield  {journal}
  {\bibinfo  {journal} {J. Appl. Phys.}\ }\textbf {\bibinfo {volume} {37}},\
  \bibinfo {pages} {1256} (\bibinfo {year} {1966})}\BibitemShut {NoStop}%
\bibitem [{\citenamefont {Ishikawa}\ \emph {et~al.}(1976)\citenamefont
  {Ishikawa}, \citenamefont {Tajima}, \citenamefont {Bloch},\ and\
  \citenamefont {Roth}}]{ishikawa}%
  \BibitemOpen
  \bibfield  {author} {\bibinfo {author} {\bibfnamefont {Y.}~\bibnamefont
  {Ishikawa}}, \bibinfo {author} {\bibfnamefont {K.}~\bibnamefont {Tajima}},
  \bibinfo {author} {\bibfnamefont {D.}~\bibnamefont {Bloch}}, \ and\ \bibinfo
  {author} {\bibfnamefont {M.}~\bibnamefont {Roth}},\ }\href {\doibase
  https://doi.org/10.1016/0038-1098(76)90057-0} {\bibfield  {journal} {\bibinfo
   {journal} {Solid State Commun.}\ }\textbf {\bibinfo {volume} {19}},\
  \bibinfo {pages} {525 } (\bibinfo {year} {1976})}\BibitemShut {NoStop}%
\bibitem [{\citenamefont {M{\"u}hlbauer}\ \emph {et~al.}(2009)\citenamefont
  {M{\"u}hlbauer}, \citenamefont {Binz}, \citenamefont {Jonietz}, \citenamefont
  {Pfleiderer}, \citenamefont {Rosch}, \citenamefont {Neubauer}, \citenamefont
  {Georgii},\ and\ \citenamefont {B{\"o}ni}}]{muhlbauer}%
  \BibitemOpen
  \bibfield  {author} {\bibinfo {author} {\bibfnamefont {S.}~\bibnamefont
  {M{\"u}hlbauer}}, \bibinfo {author} {\bibfnamefont {B.}~\bibnamefont {Binz}},
  \bibinfo {author} {\bibfnamefont {F.}~\bibnamefont {Jonietz}}, \bibinfo
  {author} {\bibfnamefont {C.}~\bibnamefont {Pfleiderer}}, \bibinfo {author}
  {\bibfnamefont {A.}~\bibnamefont {Rosch}}, \bibinfo {author} {\bibfnamefont
  {A.}~\bibnamefont {Neubauer}}, \bibinfo {author} {\bibfnamefont
  {R.}~\bibnamefont {Georgii}}, \ and\ \bibinfo {author} {\bibfnamefont
  {P.}~\bibnamefont {B{\"o}ni}},\ }\href {\doibase 10.1126/science.1166767}
  {\bibfield  {journal} {\bibinfo  {journal} {Science}\ }\textbf {\bibinfo
  {volume} {323}},\ \bibinfo {pages} {915} (\bibinfo {year}
  {2009})}\BibitemShut {NoStop}%
\bibitem [{\citenamefont {Grigoriev}\ \emph {et~al.}(2009)\citenamefont
  {Grigoriev}, \citenamefont {Dyadkin}, \citenamefont {Moskvin}, \citenamefont
  {Lamago}, \citenamefont {Wolf}, \citenamefont {Eckerlebe},\ and\
  \citenamefont {Maleyev}}]{grigoriev01}%
  \BibitemOpen
  \bibfield  {author} {\bibinfo {author} {\bibfnamefont {S.~V.}\ \bibnamefont
  {Grigoriev}}, \bibinfo {author} {\bibfnamefont {V.~A.}\ \bibnamefont
  {Dyadkin}}, \bibinfo {author} {\bibfnamefont {E.~V.}\ \bibnamefont
  {Moskvin}}, \bibinfo {author} {\bibfnamefont {D.}~\bibnamefont {Lamago}},
  \bibinfo {author} {\bibfnamefont {T.}~\bibnamefont {Wolf}}, \bibinfo {author}
  {\bibfnamefont {H.}~\bibnamefont {Eckerlebe}}, \ and\ \bibinfo {author}
  {\bibfnamefont {S.~V.}\ \bibnamefont {Maleyev}},\ }\href {\doibase
  10.1103/PhysRevB.79.144417} {\bibfield  {journal} {\bibinfo  {journal} {Phys.
  Rev. B}\ }\textbf {\bibinfo {volume} {79}},\ \bibinfo {pages} {144417}
  (\bibinfo {year} {2009})}\BibitemShut {NoStop}%
\bibitem [{\citenamefont {Bauer}\ \emph {et~al.}(2010)\citenamefont {Bauer},
  \citenamefont {Neubauer}, \citenamefont {Franz}, \citenamefont {M\"unzer},
  \citenamefont {Garst},\ and\ \citenamefont {Pfleiderer}}]{bauer2}%
  \BibitemOpen
  \bibfield  {author} {\bibinfo {author} {\bibfnamefont {A.}~\bibnamefont
  {Bauer}}, \bibinfo {author} {\bibfnamefont {A.}~\bibnamefont {Neubauer}},
  \bibinfo {author} {\bibfnamefont {C.}~\bibnamefont {Franz}}, \bibinfo
  {author} {\bibfnamefont {W.}~\bibnamefont {M\"unzer}}, \bibinfo {author}
  {\bibfnamefont {M.}~\bibnamefont {Garst}}, \ and\ \bibinfo {author}
  {\bibfnamefont {C.}~\bibnamefont {Pfleiderer}},\ }\href {\doibase
  10.1103/PhysRevB.82.064404} {\bibfield  {journal} {\bibinfo  {journal} {Phys.
  Rev. B}\ }\textbf {\bibinfo {volume} {82}},\ \bibinfo {pages} {064404}
  (\bibinfo {year} {2010})}\BibitemShut {NoStop}%
\bibitem [{\citenamefont {Dhital}\ \emph
  {et~al.}(2017{\natexlab{a}})\citenamefont {Dhital}, \citenamefont
  {DeBeer-Schmitt}, \citenamefont {Zhang}, \citenamefont {Xie}, \citenamefont
  {Young},\ and\ \citenamefont {DiTusa}}]{dhital2}%
  \BibitemOpen
  \bibfield  {author} {\bibinfo {author} {\bibfnamefont {C.}~\bibnamefont
  {Dhital}}, \bibinfo {author} {\bibfnamefont {L.}~\bibnamefont
  {DeBeer-Schmitt}}, \bibinfo {author} {\bibfnamefont {Q.}~\bibnamefont
  {Zhang}}, \bibinfo {author} {\bibfnamefont {W.}~\bibnamefont {Xie}}, \bibinfo
  {author} {\bibfnamefont {D.~P.}\ \bibnamefont {Young}}, \ and\ \bibinfo
  {author} {\bibfnamefont {J.~F.}\ \bibnamefont {DiTusa}},\ }\href {\doibase
  10.1103/PhysRevB.96.214425} {\bibfield  {journal} {\bibinfo  {journal} {Phys.
  Rev. B}\ }\textbf {\bibinfo {volume} {96}},\ \bibinfo {pages} {214425}
  (\bibinfo {year} {2017}{\natexlab{a}})}\BibitemShut {NoStop}%
\bibitem [{\citenamefont {Sivakumar}\ \emph {et~al.}(2006)\citenamefont
  {Sivakumar}, \citenamefont {Kuo},\ and\ \citenamefont {Lue}}]{sivakumar}%
  \BibitemOpen
  \bibfield  {author} {\bibinfo {author} {\bibfnamefont {K.}~\bibnamefont
  {Sivakumar}}, \bibinfo {author} {\bibfnamefont {Y.}~\bibnamefont {Kuo}}, \
  and\ \bibinfo {author} {\bibfnamefont {C.}~\bibnamefont {Lue}},\ }\href
  {\doibase https://doi.org/10.1016/j.jmmm.2006.02.066} {\bibfield  {journal}
  {\bibinfo  {journal} {J. Magn. Magn. Matter.}\ }\textbf {\bibinfo {volume}
  {304}},\ \bibinfo {pages} {e315 } (\bibinfo {year} {2006})}\BibitemShut
  {NoStop}%
\bibitem [{\citenamefont {Potapova}\ \emph {et~al.}(2012)\citenamefont
  {Potapova}, \citenamefont {Dyadkin}, \citenamefont {Moskvin}, \citenamefont
  {Eckerlebe}, \citenamefont {Menzel},\ and\ \citenamefont
  {Grigoriev}}]{nadya}%
  \BibitemOpen
  \bibfield  {author} {\bibinfo {author} {\bibfnamefont {N.}~\bibnamefont
  {Potapova}}, \bibinfo {author} {\bibfnamefont {V.}~\bibnamefont {Dyadkin}},
  \bibinfo {author} {\bibfnamefont {E.}~\bibnamefont {Moskvin}}, \bibinfo
  {author} {\bibfnamefont {H.}~\bibnamefont {Eckerlebe}}, \bibinfo {author}
  {\bibfnamefont {D.}~\bibnamefont {Menzel}}, \ and\ \bibinfo {author}
  {\bibfnamefont {S.}~\bibnamefont {Grigoriev}},\ }\href {\doibase
  10.1103/PhysRevB.86.060406} {\bibfield  {journal} {\bibinfo  {journal} {Phys.
  Rev. B}\ }\textbf {\bibinfo {volume} {86}},\ \bibinfo {pages} {060406(R)}
  (\bibinfo {year} {2012})}\BibitemShut {NoStop}%
\bibitem [{\citenamefont {Fujishiro}\ \emph {et~al.}(2019)\citenamefont
  {Fujishiro}, \citenamefont {Kanazawa}, \citenamefont {Nakajima},
  \citenamefont {Yu}, \citenamefont {Ohishi}, \citenamefont {Kawamura},
  \citenamefont {Kakurai}, \citenamefont {Arima}, \citenamefont {Mitamura},
  \citenamefont {Miyake}, \citenamefont {Akiba}, \citenamefont {Tokunaga},
  \citenamefont {Matsuo}, \citenamefont {Kindo}, \citenamefont {Koretsune},
  \citenamefont {Arita},\ and\ \citenamefont {Tokura}}]{fujishiro}%
  \BibitemOpen
  \bibfield  {author} {\bibinfo {author} {\bibfnamefont {Y.}~\bibnamefont
  {Fujishiro}}, \bibinfo {author} {\bibfnamefont {N.}~\bibnamefont {Kanazawa}},
  \bibinfo {author} {\bibfnamefont {T.}~\bibnamefont {Nakajima}}, \bibinfo
  {author} {\bibfnamefont {X.~Z.}\ \bibnamefont {Yu}}, \bibinfo {author}
  {\bibfnamefont {K.}~\bibnamefont {Ohishi}}, \bibinfo {author} {\bibfnamefont
  {Y.}~\bibnamefont {Kawamura}}, \bibinfo {author} {\bibfnamefont
  {K.}~\bibnamefont {Kakurai}}, \bibinfo {author} {\bibfnamefont
  {T.}~\bibnamefont {Arima}}, \bibinfo {author} {\bibfnamefont
  {H.}~\bibnamefont {Mitamura}}, \bibinfo {author} {\bibfnamefont
  {A.}~\bibnamefont {Miyake}}, \bibinfo {author} {\bibfnamefont
  {K.}~\bibnamefont {Akiba}}, \bibinfo {author} {\bibfnamefont
  {M.}~\bibnamefont {Tokunaga}}, \bibinfo {author} {\bibfnamefont
  {A.}~\bibnamefont {Matsuo}}, \bibinfo {author} {\bibfnamefont
  {K.}~\bibnamefont {Kindo}}, \bibinfo {author} {\bibfnamefont
  {T.}~\bibnamefont {Koretsune}}, \bibinfo {author} {\bibfnamefont
  {R.}~\bibnamefont {Arita}}, \ and\ \bibinfo {author} {\bibfnamefont
  {Y.}~\bibnamefont {Tokura}},\ }\href {\doibase 10.1038/s41467-019-08985-6}
  {\bibfield  {journal} {\bibinfo  {journal} {Nat. Commun.}\ }\textbf {\bibinfo
  {volume} {10}},\ \bibinfo {pages} {1059} (\bibinfo {year}
  {2019})}\BibitemShut {NoStop}%
\bibitem [{\citenamefont {Maleyev}(2006)}]{maleyev}%
  \BibitemOpen
  \bibfield  {author} {\bibinfo {author} {\bibfnamefont {S.~V.}\ \bibnamefont
  {Maleyev}},\ }\href {\doibase 10.1103/PhysRevB.73.174402} {\bibfield
  {journal} {\bibinfo  {journal} {Phys. Rev. B}\ }\textbf {\bibinfo {volume}
  {73}},\ \bibinfo {pages} {174402} (\bibinfo {year} {2006})}\BibitemShut
  {NoStop}%
\bibitem [{\citenamefont {Skomski}(2008)}]{ralph}%
  \BibitemOpen
  \bibfield  {author} {\bibinfo {author} {\bibfnamefont {R.}~\bibnamefont
  {Skomski}},\ }\href@noop {} {\emph {\bibinfo {title} {Simple Models of
  Magnetism}}}\ (\bibinfo  {publisher} {Oxford},\ \bibinfo {address} {New
  York},\ \bibinfo {year} {2008})\BibitemShut {NoStop}%
\bibitem [{\citenamefont {Rodriguez-Carvajal}(1993)}]{rodriguez2}%
  \BibitemOpen
  \bibfield  {author} {\bibinfo {author} {\bibfnamefont {J.}~\bibnamefont
  {Rodriguez-Carvajal}},\ }\href {\doibase
  https://doi.org/10.1016/0921-4526(93)90108-I} {\bibfield  {journal} {\bibinfo
   {journal} {Physica B}\ }\textbf {\bibinfo {volume} {192}},\ \bibinfo {pages}
  {55 } (\bibinfo {year} {1993})}\BibitemShut {NoStop}%
\bibitem [{\citenamefont {Aharoni}(1998)}]{aharoni}%
  \BibitemOpen
  \bibfield  {author} {\bibinfo {author} {\bibfnamefont {A.}~\bibnamefont
  {Aharoni}},\ }\href {\doibase 10.1063/1.367113} {\bibfield  {journal}
  {\bibinfo  {journal} {J. Appl. Phys.}\ }\textbf {\bibinfo {volume} {83}},\
  \bibinfo {pages} {3432} (\bibinfo {year} {1998})}\BibitemShut {NoStop}%
\bibitem [{\citenamefont {Bauer}\ and\ \citenamefont
  {Pfleiderer}(2016)}]{Bauer2016book}%
  \BibitemOpen
  \bibfield  {author} {\bibinfo {author} {\bibfnamefont {A.}~\bibnamefont
  {Bauer}}\ and\ \bibinfo {author} {\bibfnamefont {C.}~\bibnamefont
  {Pfleiderer}},\ }\enquote {\bibinfo {title} {Topological structures in
  ferroic materials: Domain walls, vortices and skyrmions},}\ \ (\bibinfo
  {publisher} {Springer International Publishing},\ \bibinfo {address} {Cham},\
  \bibinfo {year} {2016})\ Chap.\ \bibinfo {chapter} {Generic Aspects of
  Skyrmion Lattices in Chiral Magnets}, pp.\ \bibinfo {pages}
  {1--28}\BibitemShut {NoStop}%
\bibitem [{\citenamefont {Dhital}\ \emph
  {et~al.}(2017{\natexlab{b}})\citenamefont {Dhital}, \citenamefont {Khan},
  \citenamefont {Saghayezhian}, \citenamefont {Phelan}, \citenamefont {Young},
  \citenamefont {Jin},\ and\ \citenamefont {DiTusa}}]{dhital1}%
  \BibitemOpen
  \bibfield  {author} {\bibinfo {author} {\bibfnamefont {C.}~\bibnamefont
  {Dhital}}, \bibinfo {author} {\bibfnamefont {M.~A.}\ \bibnamefont {Khan}},
  \bibinfo {author} {\bibfnamefont {M.}~\bibnamefont {Saghayezhian}}, \bibinfo
  {author} {\bibfnamefont {W.~A.}\ \bibnamefont {Phelan}}, \bibinfo {author}
  {\bibfnamefont {D.~P.}\ \bibnamefont {Young}}, \bibinfo {author}
  {\bibfnamefont {R.~Y.}\ \bibnamefont {Jin}}, \ and\ \bibinfo {author}
  {\bibfnamefont {J.~F.}\ \bibnamefont {DiTusa}},\ }\href {\doibase
  10.1103/PhysRevB.95.024407} {\bibfield  {journal} {\bibinfo  {journal} {Phys.
  Rev. B}\ }\textbf {\bibinfo {volume} {95}},\ \bibinfo {pages} {024407}
  (\bibinfo {year} {2017}{\natexlab{b}})}\BibitemShut {NoStop}%
\bibitem [{\citenamefont {Grigoriev}\ \emph {et~al.}(2006)\citenamefont
  {Grigoriev}, \citenamefont {Maleyev}, \citenamefont {Okorokov}, \citenamefont
  {Chetverikov}, \citenamefont {B\"oni}, \citenamefont {Georgii}, \citenamefont
  {Lamago}, \citenamefont {Eckerlebe},\ and\ \citenamefont
  {Pranzas}}]{gregorievmnsi}%
  \BibitemOpen
  \bibfield  {author} {\bibinfo {author} {\bibfnamefont {S.~V.}\ \bibnamefont
  {Grigoriev}}, \bibinfo {author} {\bibfnamefont {S.~V.}\ \bibnamefont
  {Maleyev}}, \bibinfo {author} {\bibfnamefont {A.~I.}\ \bibnamefont
  {Okorokov}}, \bibinfo {author} {\bibfnamefont {Y.~O.}\ \bibnamefont
  {Chetverikov}}, \bibinfo {author} {\bibfnamefont {P.}~\bibnamefont {B\"oni}},
  \bibinfo {author} {\bibfnamefont {R.}~\bibnamefont {Georgii}}, \bibinfo
  {author} {\bibfnamefont {D.}~\bibnamefont {Lamago}}, \bibinfo {author}
  {\bibfnamefont {H.}~\bibnamefont {Eckerlebe}}, \ and\ \bibinfo {author}
  {\bibfnamefont {K.}~\bibnamefont {Pranzas}},\ }\href {\doibase
  10.1103/PhysRevB.74.214414} {\bibfield  {journal} {\bibinfo  {journal} {Phys.
  Rev. B}\ }\textbf {\bibinfo {volume} {74}},\ \bibinfo {pages} {214414}
  (\bibinfo {year} {2006})}\BibitemShut {NoStop}%
\bibitem [{\citenamefont {Chizhikov}\ and\ \citenamefont
  {Dmitrienko}(2012)}]{chizhikov}%
  \BibitemOpen
  \bibfield  {author} {\bibinfo {author} {\bibfnamefont {V.~A.}\ \bibnamefont
  {Chizhikov}}\ and\ \bibinfo {author} {\bibfnamefont {V.~E.}\ \bibnamefont
  {Dmitrienko}},\ }\href {\doibase 10.1103/PhysRevB.85.014421} {\bibfield
  {journal} {\bibinfo  {journal} {Phys. Rev. B}\ }\textbf {\bibinfo {volume}
  {85}},\ \bibinfo {pages} {014421} (\bibinfo {year} {2012})}\BibitemShut
  {NoStop}%
\bibitem [{\citenamefont {Ishikawa}\ \emph {et~al.}(1977)\citenamefont
  {Ishikawa}, \citenamefont {Shirane}, \citenamefont {Tarvin},\ and\
  \citenamefont {Kohgi}}]{ishikawa2}%
  \BibitemOpen
  \bibfield  {author} {\bibinfo {author} {\bibfnamefont {Y.}~\bibnamefont
  {Ishikawa}}, \bibinfo {author} {\bibfnamefont {G.}~\bibnamefont {Shirane}},
  \bibinfo {author} {\bibfnamefont {J.~A.}\ \bibnamefont {Tarvin}}, \ and\
  \bibinfo {author} {\bibfnamefont {M.}~\bibnamefont {Kohgi}},\ }\href
  {\doibase 10.1103/PhysRevB.16.4956} {\bibfield  {journal} {\bibinfo
  {journal} {Phys. Rev. B}\ }\textbf {\bibinfo {volume} {16}},\ \bibinfo
  {pages} {4956} (\bibinfo {year} {1977})}\BibitemShut {NoStop}%
\bibitem [{\citenamefont {Grigoriev}\ \emph {et~al.}(2015)\citenamefont
  {Grigoriev}, \citenamefont {Sukhanov}, \citenamefont {Altynbaev},
  \citenamefont {Siegfried}, \citenamefont {Heinemann}, \citenamefont {Kizhe},\
  and\ \citenamefont {Maleyev}}]{grigoriev02}%
  \BibitemOpen
  \bibfield  {author} {\bibinfo {author} {\bibfnamefont {S.~V.}\ \bibnamefont
  {Grigoriev}}, \bibinfo {author} {\bibfnamefont {A.~S.}\ \bibnamefont
  {Sukhanov}}, \bibinfo {author} {\bibfnamefont {E.~V.}\ \bibnamefont
  {Altynbaev}}, \bibinfo {author} {\bibfnamefont {S.-A.}\ \bibnamefont
  {Siegfried}}, \bibinfo {author} {\bibfnamefont {A.}~\bibnamefont
  {Heinemann}}, \bibinfo {author} {\bibfnamefont {P.}~\bibnamefont {Kizhe}}, \
  and\ \bibinfo {author} {\bibfnamefont {S.~V.}\ \bibnamefont {Maleyev}},\
  }\href {\doibase 10.1103/PhysRevB.92.220415} {\bibfield  {journal} {\bibinfo
  {journal} {Phys. Rev. B}\ }\textbf {\bibinfo {volume} {92}},\ \bibinfo
  {pages} {220415(R)} (\bibinfo {year} {2015})}\BibitemShut {NoStop}%
\bibitem [{\citenamefont {Pappas}\ \emph {et~al.}(2011)\citenamefont {Pappas},
  \citenamefont {Leli\`evre-Berna}, \citenamefont {Bentley}, \citenamefont
  {Falus}, \citenamefont {Fouquet},\ and\ \citenamefont {Farago}}]{papas}%
  \BibitemOpen
  \bibfield  {author} {\bibinfo {author} {\bibfnamefont {C.}~\bibnamefont
  {Pappas}}, \bibinfo {author} {\bibfnamefont {E.}~\bibnamefont
  {Leli\`evre-Berna}}, \bibinfo {author} {\bibfnamefont {P.}~\bibnamefont
  {Bentley}}, \bibinfo {author} {\bibfnamefont {P.}~\bibnamefont {Falus}},
  \bibinfo {author} {\bibfnamefont {P.}~\bibnamefont {Fouquet}}, \ and\
  \bibinfo {author} {\bibfnamefont {B.}~\bibnamefont {Farago}},\ }\href
  {\doibase 10.1103/PhysRevB.83.224405} {\bibfield  {journal} {\bibinfo
  {journal} {Phys. Rev. B}\ }\textbf {\bibinfo {volume} {83}},\ \bibinfo
  {pages} {224405} (\bibinfo {year} {2011})}\BibitemShut {NoStop}%
\end{thebibliography}


\end{document}